\documentstyle[twoside,fleqn,espcrc2]{article}
\def\bar#1{\overline{#1}}

\title{Chiral Perturbation Theory and Weak Matrix Elements}

\author{Stephen R. Sharpe\address{Physics Department, 
University of Washington, Seattle, WA 98195, USA}%
\thanks{Plenary talk at LATTICE 96. An abridged version will appear
in the proceedings. Preprint UW/PT-96-16. Email: sharpe@phys.washington.edu}
}
       
\begin{document}
\input{psfig}

\begin{abstract}
%
I describe recent developments in quenched chiral perturbation theory (QChPT) 
and the status of weak matrix elements involving light quarks. I illustrate 
how, with improved statistical errors, and with calculations of the masses of 
baryons containing non-degenerate quarks, there is now a clear need for 
extrapolations of higher order than linear in the quark mass. I describe how 
QChPT makes predictions for the functional forms to use in such extrapolations,
and emphasize the distinction between contributions coming from chiral loops 
which are similar to those present in unquenched theories, and those from 
$\eta'$ loops which are pure quenched artifacts. I describe a fit to the 
baryon masses using the predictions of QChPT. I give a status report on the 
numerical evidence for $\eta'$ loops, concluding that they are likely present,
and are characterized by a coupling $\delta=0.1-0.2$. I use the difference 
between chiral loops in QCD and quenched QCD to estimate the quenching errors 
in a variety of quantities. I then turn to results for matrix elements, 
largely from quenched simulations. Results for quenched decay constants 
cannot yet be reliably extrapolated to the continuum limit. By contrast, 
new results for $B_K$ suggest a continuum, {\em quenched} value of
$B_K({\rm NDR}, 2\,{\rm GeV}) = 0.5977  \pm 0.0064 \pm 0.0166$,
based on a quadratic extrapolation in $a$. The theoretical basis for using 
a quadratic extrapolation has been confirmed. For the first time there is 
significant evidence that unquenching changes $B_K$, and my estimate for 
the value in QCD is $B_K({\rm NDR}, 2\,{\rm GeV}) = 0.66 \pm 0.02 \pm 0.11$.
Here the second error is a conservative estimate of the systematic error due 
to uncertainties in the effect of quenching. A less conservative viewpoint 
reduces $0.11$ to $0.03$. Non-perturbative renormalization has been 
successfully applied to four-fermion operators, and has been used to 
calculate $B_K$ with Wilson fermions. The results are consistent with those for
staggered fermions. Results for other matrix elements are mentioned briefly.
\end{abstract}

\maketitle

\section{INTRODUCTION}
This talk reviews the status of two distinct though related subjects:
our understanding of chiral extrapolations, 
and results for weak matrix elements involving light (i.e. $u$, $d$
and $s$) quarks.
The major connection between these subjects
is that understanding chiral extrapolations allows
us to reduce, or at least estimate, the errors in matrix elements.
Indeed, in a number of matrix elements, the dominant errors are those
due to chiral extrapolation and to the use of the quenched approximation.
I will argue that an understanding of chiral extrapolations gives us
a handle on both of these errors.

While understanding errors in detail is a sign of a maturing field,
we are ultimately interested in the results for the matrix elements 
themselves.
The phenomenological implications of these results were emphasized
in previous reviews \cite{martinelli94,soni95}.
Here I only note that it is important to calculate matrix elements 
for which we know the answer, e.g. $f_\pi/M_\rho$ and $f_K/f_\pi$,
in order to convince ourselves, and others, that our predictions for
unknown matrix elements are reliable.
The reliability of a result for $f_D$, for example, will be gauged 
in part by how well we can calculate $f_\pi$.
And it would be a real coup if we were able to show in detail that
QCD indeed explains the $\Delta I=1/2$ rule in $K\to\pi\pi$ decays.

But to me the most interesting part of the enterprise is the possibility
of using the lattice results to calculate quantities which allow us to
test the Standard Model. In this category, the light-quark matrix element 
with which we have had the most success is $B_K$.
The lattice result is already used by phenomenological analyses which
attempt to determine the CP violation in the CKM matrix
from the experimental number for $\epsilon$.
I describe below the latest twists in the saga of the lattice
result for $B_K$.
What we would like to do is extend this success to the raft of
$B$-parameters which are needed to predict $\epsilon'/\epsilon$.
There has been some progress this year on the
contributions from electromagnetic penguins,
but we have made no headway towards calculating strong penguins.
I note parenthetically that another input into the prediction of
$\epsilon'/\epsilon$ is the strange quark mass.
The recent work of Gupta and Bhattacharya \cite{rajanmq96} 
and the Fermilab group \cite{mackenzie96} 
suggest a value of $m_s$ considerably smaller than the
accepted phenomenological estimates, which will substantially
increase the prediction for $\epsilon'/\epsilon$.

Much of the preceding could have been written in 1989, when I
gave the review talk on weak matrix elements \cite{sharpe89}.
How has the field progressed since then?
I see considerable advances in two areas. 
First, the entire field of weak matrix elements involving heavy-light
hadrons has blossomed. 1989 was early days in our calculation of the
simplest such matrix elements, $f_D$ and $f_B$. In 1996 a plethora
of quantities are being calculated, and the subject deserves its own
plenary talk \cite{flynn96}.
Second, while the subject of my talks then and now is similar, there has
been enormous progress in understanding and reducing systematic errors.
Thus, whereas in 1989 I noted the possibility of using chiral loops 
to estimate quenching errors, we now have a technology (quenched chiral
perturbation theory---QChPT) which allows us to make these estimates.
We have learned that the quenched approximation (QQCD) is most probably
singular in the chiral limit, and there is a growing body of numerical
evidence showing this, although the case is not closed.
The reduction in statistical errors has allowed us to
go beyond simple linear extrapolations in light quark masses, 
and thus begin to test the predictions of QChPT.
The increase in computer power has allowed us to study systematically the 
dependence of matrix elements on the lattice spacing, $a$.
We have learned how to get more reliable estimates
using lattice perturbation theory \cite{lepagemackenzie}.
And, finally, we have begun to use non-perturbative matching of lattice 
and continuum operators, as discussed here by Rossi \cite{rossi96}.

The body of this talk is divided into two parts. In the first,
Secs. \ref{sec:whychi}-\ref{sec:querr}, I focus on chiral extrapolations:
why we need them, how we calculate their expected form,
the evidence for chiral loops, and how we can use them to estimate
quenching errors.
In the second part, Secs. \ref{sec:decayc}-\ref{sec:otherme},
I give an update on results for weak matrix elements.
I discuss $f_\pi/M_\rho$, $f_K/f_\pi$, $B_K$ and a few related $B$-parameters.
Results for structure functions have been reviewed here
by G\"ockeler \cite{gockeler96}.
There has been little progress on semi-leptonic form factors, 
nor on flavor singlet matrix elements and scattering lengths, 
since last years talks by Simone \cite{simone95} and Okawa \cite{okawa95}.

\section{WHY DO WE NEED QChPT?}
\label{sec:whychi}
\begin{figure}[tb]
\vspace{-0.1truein}
\centerline{\psfig{file=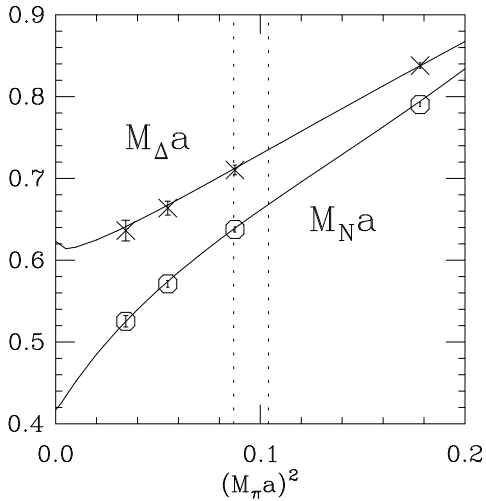,height=3.0truein}}
\vspace{-0.6truein}
\caption{$M_N$ and $M_\Delta$ versus $M_\pi^2$,
with quenched Wilson fermions, at $\beta=6$, on $32^3\times64$
lattices. The vertical band is the range of estimates for $m_s$.}
\vspace{-0.2truein}
\label{fig:NDelta}
\end{figure}

Until recently linear chiral extrapolations 
(i.e. $\alpha + \beta m_q$) have sufficed for most quantities.
This is no longer true.
This change has come about for two reasons. First, smaller statistical
errors (and to some extent the use of a larger range of quark masses)
have exposed the inadequacies of linear fits, as already stressed here
by Gottlieb \cite{gottlieb96}.
An example, taken from Ref. \cite{ourspect96}
is shown in Fig. \ref{fig:NDelta}.
The range of quark masses is $m_s/3- 2 m_s$,
typical of that in present calculations.
While $M_\Delta$ is adequately fit by a straight line,
there is definite, though small, curvature in $M_N$.
The curves are the result of a fit using QChPT \cite{sharpebary}, 
to which I return below.

\begin{figure}[tb]
\vspace{-0.1truein}
\centerline{\psfig{file=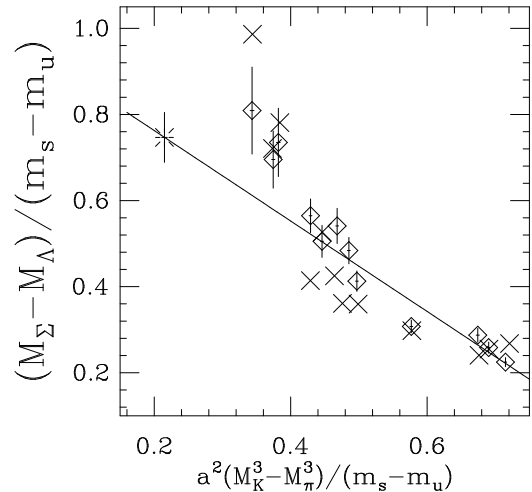,height=3.0truein}}
\vspace{-0.6truein}
\caption{Results for $(M_\Sigma-M_\Lambda)/(m_s-m_u)$ (diamonds).
The ``burst'' is the physical result using the linear fit shown. 
Crosses are from a global chiral fit.}
\vspace{-0.2truein}
\label{fig:sigdel}
\end{figure}

The second demonstration of the failure of linear extrapolations
has come from studying the octet baryon mass splittings \cite{ourspect96}.
The new feature here is the consideration of baryons
composed of non-degenerate quarks.
I show the results for $(M_\Sigma-M_\Lambda)/(m_s-m_u)$ (with $m_u=m_d$)
in Fig. \ref{fig:sigdel}. 
If baryons masses were linear functions of $m_s$ and $m_u$
then the data would lie on a horizontal line.
Instead the results vary by a factor of four.
This is a glaring deviation from linear behavior,
in contrast to the subtle effect shown in Fig. \ref{fig:NDelta}.

Once there is evidence of non-linearity, we need to know
the appropriate functional form with which to extrapolate to
the chiral limit. This is where (Q)ChPT comes in.
In the second example, the prediction is \cite{LS,sharpebary}
\begin{eqnarray}
\lefteqn{{M_\Sigma - M_\Lambda \over m_s-m_u} \approx} \nonumber\\
&& - {8 D\over3} 
+ d_1 {M_K^3-M_\pi^3 \over m_s-m_u}
 + d'_1 {M_{ss}^3-M_\pi^3 \over m_s-m_u}   \nonumber \\
&& + e_1 (m_u + m_d + m_s) + e'_1 (m_u + m_s) \,, 
\label{eq:chptsiglam}
\end{eqnarray}
where $M_{ss}$ is the mass of the quenched $\bar ss$ meson.\footnote{%
I have omitted terms which, though more singular in the chiral limit,
are expected to be numerically small for the range of $m_q$ under study.}
QChPT also implies that $|d'_1| < |d_1|$, which is why in
Fig. \ref{fig:sigdel} I have plotted the data versus
$(M_K^3-M_\pi^3)/(m_s-m_u)$. The good news is that the points do lie
approximately on a single curve---which would not be true for a poor
choice of y-axis. The bad news is that the fit to the 
the $d_1$ term and a constant is not that good.

This example illustrates the benefits which can accrue if one knows
the chiral expansion of the quantity under study.
First, the data collapses onto a single curve, allowing
an extrapolation to the physical quark masses.
And, second, the theoretical input reduces the length of the
extrapolation. 
In Fig. \ref{fig:sigdel}, for example, to reach the physical point requires an
extrapolation by less than a factor of 2.
This is much smaller than the ratio of the lightest quark mass to
the physical value of $(m_u+m_d)/2$---a factor of roughly 8.

In fact, in the present example,
the data are not yet accurate enough to distinguish between 
the $d$ and $e$ terms in Eq. \ref{eq:chptsiglam}.
A global fit of the QChPT prediction to this and other mass differences 
(the results of which are shown in the Figure)
implies that both types of term are present \cite{sharpebary}.

\section{
CHIRAL PERTURBATION THEORY}
\label{sec:qchpt}

This brings me to a summary of QChPT.
In QCD, chiral perturbation theory predicts the form of the chiral expansion
for quantities involving one or more light quarks.
The expansion involves terms analytic in $m_q$ and the external momenta,
and non-analytic terms due to pion loops.\footnote{``Pion'' here refers to
any of the pseudo-Goldstone bosons.}
The analytic terms are restricted in form, 
though not in magnitude, by chiral symmetry,
while the non-analytic terms are completely predicted given the 
analytic terms.

The same type of expansions can be developed in quenched QCD using QChPT.
The method was worked out in Refs. \cite{morel,sharpestcoup,BGI,sharpechbk},
with the theoretically best motivated formulation 
being that of Bernard and Golterman \cite{BGI}.
Their method 
gives a precise meaning to the quark-flow diagrams which I use below.
They have extended it also to partially quenched theories 
(those having both valence and sea quarks but with different masses)
\cite{BGPQ}.
Results are available for pion properties and the condensate \cite{BGI},
$B_K$ and related matrix elements \cite{sharpechbk},
$f_B$, $B_B$ and the Isgur-Wise function \cite{booth,zhang},
baryon masses \cite{sharpebary,LS},
and scattering lengths \cite{BGscat}.
I will not describe technical details, but rather focus on
the aims and major conclusions of the approach.
For a more technical review see Ref. \cite{maartench}.

As I see it, the major aims of QChPT are these. \\
$\bullet\ $
To predict the form of the chiral expansion, which can then be
used to fit and extrapolate the data. This was the approach taken above
for the baryon masses. \\
$\bullet\ $
To estimate the size of the quenching error by comparing the 
contribution of the pion loops in QCD and QQCD,
using the coefficients of the chiral fits in QCD (from phenomenology)
and in QQCD (from a fit to the lattice data).
I return to such estimates in Sec. \ref{sec:querr}.

\begin{figure}[tb]
\centerline{\psfig{file=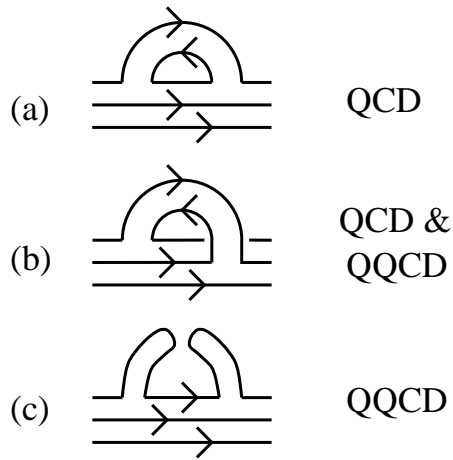,height=2.4truein}}
\vspace{-0.3truein}
\caption{Quark flow diagrams for $M_{\rm bary}$.}
\vspace{-0.2truein}
\label{fig:quarkloops}
\end{figure}

I begin by describing the form of the chiral expansions,
and in particular how they are affected by quenching.
The largest changes are to the non-analytic terms,
i.e. those due to pion loops.
There are two distinct effects.\\
(1) Quenching removes loops which, at the underlying quark level,
require an internal loop.
This is illustrated for baryon masses in Fig. \ref{fig:quarkloops}.
Diagrams of types (a) and (b) contribute in QCD,
but only type (b) occurs in QQCD.
These loops give rise to $M_\pi^3$ terms in the chiral expansions.
Thus for baryon masses, quenching only changes the coefficient of
these terms. In other quantities, e.g. $f_\pi$, they are
removed entirely. \\
(2) Quenching introduces artifacts due to $\eta'$ loops---as
in Fig. \ref{fig:quarkloops}(c).
These are chiral loops because the $\eta'$ remains light in QQCD.
Their strength is determined by the size of the ``hairpin'' vertex, 
and is parameterized by $\delta$ and $\alpha_\Phi$ 
(defined in Sec. \ref{sec:chevidence} below).

The first effect of quenching is of greater practical importance, 
and dominates most estimates of quenching errors.
It does not change the form of the chiral expansion, 
only the size of the terms. 

The second effect leads to new terms in the chiral expansion,
some of which are singular in the chiral limit. 
If these terms are large, 
then one can be sure that quenching errors are large. 
One wants to work at large enough quark mass so that these new terms are
numerically small. In practice this means keeping $m_q$ above 
about $m_s/4-m_s/3$.

I illustrate these general comments using the results of 
Labrenz and I for baryon masses \cite{LS}.
The form of the chiral expansion in QCD is\footnote{%
There are also $M_\pi^4 \log M_\pi$ terms in both QCD and QQCD,
the coefficients of which have not yet been calculated in QQCD.
For the limited range of quark masses used in simulations,
I expect that these terms can be adequately represented by the
$M_\pi^4$ terms, whose coefficients are unknown parameters.}
\begin{equation}
M_{\rm bary} = M_0 + c_2 M_\pi^2 + c_3 M_\pi^3 + c_4 M_\pi^4 + \dots 
\end{equation}
with $c_3$ predicted in terms of $g_{\pi NN}$ and $f_\pi$.
In QQCD 
\begin{eqnarray}
\lefteqn{
M_{\rm bary}^Q = M_0^Q + c_2^Q M_\pi^2 + c_3^Q M_\pi^3 + c_4^Q M_\pi^4 + }
\nonumber \\
&& \delta \left( c_1^Q M_\pi + c_2^Q M_\pi^2 \log M_\pi\right) + \alpha_\Phi
\tilde c_3^Q M_\pi^3 + \dots
\label{eq:baryqchpt}
\end{eqnarray}
The first line has the same form as in QCD,
although the constants multiplying the analytic terms in the two theories
are unrelated.
$c_3^Q$ is predicted to be non-vanishing, though different from $c_3$.
The second line is the contribution of $\eta'$ loops and is a quenched
artifact. 
Note that it is the dominant correction in the chiral limit.

In order to test QChPT in more detail,
I have attempted to fit the expressions outlined above
to the octet and decuplet baryon masses from Ref. \cite{ourspect96}.
There are 48 masses, to be fit in terms of 19 underlying
parameters: the octet and decuplet masses in the chiral limit,
3 constants of the form $c_2^Q$, 6 of the form $c_4^Q$,
6 pion-nucleon couplings, $\delta$ and $\alpha_\Phi$.
I have found a reasonable description of the data with $\delta\approx0.1$,
but the errors are too large to pin down the values 
of all the constants \cite{sharpebary}.
Examples of the fit are shown in Figs. \ref{fig:NDelta} and \ref{fig:sigdel}.

\begin{figure}[tb]
\vspace{-0.1truein}
\centerline{\psfig{file=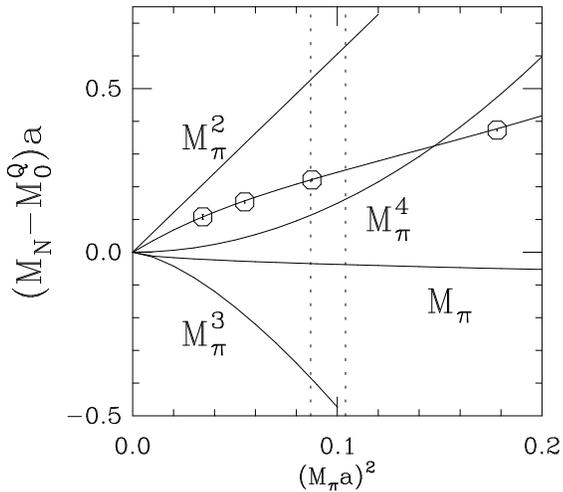,height=3.0truein}}
\vspace{-0.6truein}
\caption{Contributions to $M_N - M_0^Q$ in the global chiral fit.
All quantities in lattice units. 
Vertical lines indicate range of estimates of $m_s$.}
\vspace{-0.2truein}
\label{fig:mnchfit}
\end{figure}

Although this is a first, rather crude, attempt at such a fit, 
several important lessons emerge.\\
(1) For present quark masses, one needs several terms in the chiral
expansion to fit the data.
This in turn requires that one have high statistics results for a
number of light quark masses. \\
(2) One must check that the ``fit'' is not resulting from
large cancellations between different orders.
The situation for my fit is illustrated by Fig. \ref{fig:mnchfit},
where I show the different contributions to $M_N$.
Note that the relative size of the terms
is not determined by $M_N$, but rather by the fit as a whole.
The most important point is that the $M_\pi^4$ terms are considerably smaller
than those of $O(M_\pi^2)$ up to at least $m_q= m_s$.
The $M_\pi^3$ terms are part of a different series and need not be
smaller than those of $O(M_\pi^2)$.
Similarly the $M_\pi$ terms are the leading quenched artifact, and
should not be compared to the other terms.
Thus the convergence is acceptable for $m_q < m_s$, though it is dubious for
the highest mass point. \\
(3) The artifacts (in particular the $\delta M_\pi$ terms)
can lead to unusual behavior at small $M_\pi$,
as illustrated in the fit to $M_\Delta$ (Fig. \ref{fig:NDelta}). \\
(4) Since the ``$\delta$-terms'' are artifacts of quenching,
and their relative contribution increases as $M_\pi\to0$,
it makes more sense phenomenologically {\em to extrapolate 
without including them}. In other words, a better estimate of the
unquenched value for $M_\Delta$ in the chiral limit
can probably be obtained simply using a linear extrapolation in $M_\pi^2$.
This is, however, a complicated issue which needs more thought.\\
(5) The output of the fit includes pion-nucleon couplings whose
values should be compared to more direct determinations. \\
(6) Finally, the fact that a fit can be found at all
gives me confidence to stick my neck out and
proceed with the estimates of quenching errors in baryon masses. 
It should be noted, however, that a fit involving only analytic terms,
including up to $M_\pi^6$, can probably not be ruled out.

What of quantities other than baryon masses?
In Sec. \ref{subsec:bkchfit} I discuss fits to $B_K$,
another quantity in which chiral loops survive quenching.
The data is consistent with the non-analytic term predicted by QChPT.
Good data also exists for $M_\rho$.
It shows curvature,
but is consistent with either cubic or quartic terms\cite{sloan96}.
What do we expect from QChPT?
In QCD the chiral expansion for $M_\rho$
has the same form as for baryon masses \cite{rhochpt}. 
The QChPT theory calculation has not been done,
but it is simple to see that form will be as for baryons,
Eq.~\ref{eq:baryqchpt}, {\em except that $c_3^Q=0$}. 
Thus an $M_\pi^3$ term is entirely a quenched
artifact---and a potential window on $\alpha_\Phi$.

What of quantities involving pions, for which there is very good data?
For the most part, quenching simply
removes the non-analytic terms of QCD and replaces them with artifacts
proportional to $\delta$. 
The search for these is the subject of the next section.

\section{EVIDENCE FOR $\eta'$ LOOPS}
\label{sec:chevidence}

The credibility of QChPT rests in part on the observation of
the singularities predicted in the chiral limit.
If such quenched artifacts are present, 
then we need to study them if only to know at
what quark masses to work in order to avoid them!
What follows in this section is an update of the 1994 review of Gupta
\cite{gupta94}.

The most direct way of measuring $\delta$ is from the $\eta'$ correlator.
If the quarks are degenerate, then the part of the quenched chiral
Lagrangian bilinear in the $\eta'$ is \cite{BGI,sharpechbk}
\begin{eqnarray}
2 {\cal L}_{\eta'} &=& \partial_\mu \eta' \partial^\mu \eta' - M_\pi^2 \eta'^2
\\
&&+  {(N_f/3)} \left( \alpha_\Phi \partial_\mu \eta' \partial^\mu \eta' - 
m_0^2 \eta'^2 \right) \,.
\end{eqnarray}
$\delta$ is defined by $\delta = m_0^2/(48 \pi^2 f_\pi^2)$.
In QQCD the terms in the second line must be treated as vertices, and 
cannot be iterated. They contribute to the disconnected part of the 
$\eta'$ correlator, whereas the first line determines the connected part.
Thus the $\eta'$ is degenerate with the pion, but has additional vertices.
To study this, various groups have looked at the ratio of the
disconnected to connected parts, at $\vec p=0$, whose predicted form 
at long times is
\begin{equation}
R(t) = t (N_f/3) (m_0^2 - \alpha_\Phi M_\pi^2)/(2 M_\pi) \,.
\end{equation}
I collect the results in Table \ref{tab:etapres}, converting 
$m_0^2$ into $\delta$ using $a$ determined from $m_\rho$.

\begin{table}[tb]
\caption{Results from the quenched $\eta'$ two point function. 
$W$ and $S$ denote Wilson and staggered fermions.}
\label{tab:etapres}
\begin{tabular}{ccccc}
\hline
Ref.		& Yr. &$\delta$&$\alpha_\Phi$	&$\beta$ W/S \\
\hline
JLQCD\cite{kuramashi94} &94&$0.14 (01)$	& $0.7$		& $5.7$ W \\
OSU\cite{kilcup95}&95& $0.27 (10)$	& $0.6$		& $6.0$ S \\
Rome\cite{masetti96}&96& $\approx 0.15$ &		& $5.7$ W \\
OSU\cite{venkat96}& 96&$0.19(05)$	& $0.6$	& $6.0$ S \\
FNAL\cite{thacker96}	&96& $< 0.02$ 	& $>0$	& $5.7$ W \\
\hline
\end{tabular}
\vspace{-0.2truein}
\end{table}

\begin{figure}[tb]
\vspace{-0.3truein}
\centerline{\psfig{file=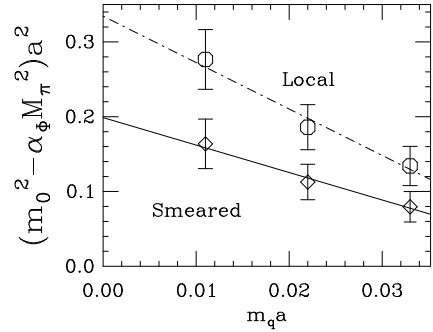,height=2.4truein}}
\vspace{-0.5truein}
\caption{$a^2(m_0^2-\alpha_\Phi M_\pi^2)$ from the OSU group.}
\vspace{-0.2truein}
\label{fig:osuR}
\end{figure}

All groups except Ref. \cite{thacker96} report a non-zero value for $\delta$
in the range $0.1-0.3$.
What they actually measure, as illustrated in Fig. \ref{fig:osuR},
is the combination $m_0^2-\alpha_\Phi M_\pi^2$,
which they then extrapolate to $M_\pi=0$. 
I have extracted the results for $\alpha_\Phi$ from such plots.
As the figure shows, there is a considerable cancellation between the
$m_0$ and $\alpha_\Phi$ terms at the largest quark masses, which
correspond to $M_\pi \approx 0.8\,$GeV.
This may explain why Ref. \cite{thacker96} 
does not see a signal for $\delta$.

Clearly, further work is needed to
sort out the differences between the various groups. 
As emphasized by Thacker \cite{thacker96}, this is mainly an issue of
understanding systematic errors.
In particular, contamination from excited states leads to an 
apparent linear rise of $R(t)$, and thus to an overestimate of $m_0^2$.
Indeed, the difference between the OSU results last year and this
is the use of smeared sources to reduce such contamination.
This leads to a smaller $\delta$, as shown in Fig. \ref{fig:osuR}.
Ref. \cite{thacker96} also find that $\delta$ decreases as the volume
is increased. 

I want to mention also that the $\eta'$ correlator has also been studied 
in partially quenched theories, with $N_f=-6, -4, -2$, 
\cite{masetti96} and $N_f=2, 4$ \cite{venkat96}.
The former work is part of the ``bermion'' program which aims to
extrapolate from negative to positive $N_f$.
For any non-zero $N_f$ the analysis is different than in the quenched
theory, because the hairpin vertices do iterate, and
lead to a shift in the $\eta'$ mass. Indeed $m_{\eta'}$ is reduced (increased)
for $N_f<0$ ($>0$), and both changes are observed!
This gives me more confidence in the results of these groups at $N_f=0$.
The bottom line appears to be
that there is relatively little dependence of $m_0^2$ on $N_f$.

Other ways of obtaining $\delta$ rely on loop effects,
such as that in Fig. \ref{fig:quarkloops}(c).
For quenched pion masses $\eta'$ loops lead to terms which are
singular in the chiral limit \cite{BGI}\footnote{%
The $\alpha_\Phi$ vertex leads to terms proportional
to $M_\pi^2 \log M_\pi$ which are not singular in the chiral limit,
and can be represented approximately by analytic terms.}
\begin{eqnarray}
\lefteqn{{M_{12}^2 \over m_1 + m_2} = \mu^Q \left[
1 - \delta \left\{\log{\widetilde M_{11}^2\over \Lambda^2} \right. \right.}
\nonumber \\
&& \left.\left.
+ {\widetilde M_{22}^2 \over \widetilde M_{22}^2 -\widetilde M_{11}^2 }
\log{\widetilde M_{22}^2 \over \widetilde M_{11}^2} \right\} 
+ c_2 (m_1 + m_2) \right] 
\label{eq:mpichpt}
\end{eqnarray}
Here $M_{ij}$ is the mass of pion
composed of a quark of mass $m_i$ and antiquark of mass $m_j$,
$\Lambda$ is an unknown scale, and $c_2$ an unknown constant.
The tilde is relevant only to staggered fermions, and indicates that it
is the mass of the flavor singlet pion, 
and not of the lattice pseudo-Goldstone pion, which appears.
This is important because, at finite lattice spacing,
$\tilde M_{ii}$ does not vanish in the chiral limit,
so there is no true singularity.

\begin{figure}[tb]
\vspace{-0.1truein}
\centerline{\psfig{file=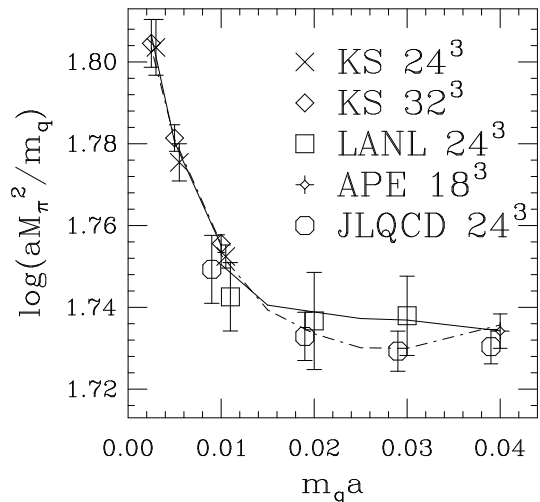,height=3.0truein}}
\vspace{-0.6truein}
\caption{Chiral fit to $\log(M_\pi^2/m_q)$ at $\beta=6$. 
Some points have been offset for clarity.}
\vspace{-0.2truein}
\label{fig:mpibymq}
\end{figure}

In his 1994 review, Gupta fit the world's data for 
staggered fermions at $\beta=6$ having $m_1=m_2$. 
I have updated his plot, including new JLQCD data \cite{yoshie96},
in Fig. \ref{fig:mpibymq}. To set the scale, note that $m_s a \approx 0.024$.
The dashed line is Gupta's fit to Eq. \ref{eq:mpichpt},  
giving $\delta=0.085$, 
while the solid line includes also an $m_q^2$ term, and gives $\delta=0.13$.
These non-zero values were driven by the results from Kim and Sinclair (KS),
who use quark masses as low as $0.1 m_s$ \cite{kimsinclair},
but they are now supported by the JLQCD results.

Last year, Mawhinney proposed an alternative
explanation for the increase visible at small $m_q$ \cite{mawhinney95},
namely an offset in the intercept of $M_\pi^2$
\begin{equation}
M_{12}^2 = c_0 + \mu^Q (m_1 + m_2) + \dots
\label{eq:mpimawh}
\end{equation}
In his model, $c_0$ is proportional to the 
minimum eigenvalue of the Dirac operator, and thus falls as $1/V$.
This model explains the detailed structure of his results for
$M_\pi^2$ and $\langle \bar\psi\psi \rangle$ at $\beta=5.7$.
It also describes the data of KS, {\em except for volume dependence of $c_0$}.
As the Fig. \ref{fig:mpibymq} shows, 
the results of KS from $24^3$ and $32^3$ lattices are consistent, 
whereas in Mawhinney's model the rise at small $m_q$
should be reduced in amplitude by $0.4$ on the larger lattice.

The fit in Fig. \ref{fig:mpibymq} is for pions composed of degenerate quarks.
One can further test QChPT by noting that Eq. \ref{eq:mpichpt} is
not simply a function of the average quark mass---there is a predicted
dependence on $m_1-m_2$. In Mawhinney's model, this dependence would
presumably enter only through a $(m_1-m_2)^2$ term, and thus would be
a weaker effect.
JLQCD have extensive data from the range $\beta=5.7-6.4$,
with both $m_1=m_2$ and $m_1\ne m_2$.
They have fit to Eq. \ref{eq:mpichpt}, and thus obtained
$\delta$ as a function of $\beta$.
They find reasonable fits, with $\delta\approx 0.06$ for most $\beta$.
I have several comments on their fits.
First, they have used $M_{ii}$ rather than $\widetilde M_{ii}$,
which leads to an underestimate of $\delta$, particularly at the
smaller values of $\beta$.
Second, their results for the constants, particularly $c_2$, vary rapidly with
$\beta$. One would expect that all dimensionless parameters in the fit 
(which are no less physical than, say, $f_\pi/M_\rho$) 
should vary smoothly and slowly with $\beta$. 
This suggests to me that terms of $O(m_q^2)$ may be needed.
Finally, it would be interesting to attempt a fit to the JLQCD data
along the lines suggested by Mawhinney, but including a $(m_1-m_2)^2$ term.

Clearly more work is needed to establish convincingly that there
are chiral singularities in $M_\pi$.
One should keep in mind that the effects are small,
$\sim 5\%$ at the lightest $m_q$,
so it is impressive that we can study them at all.
Let me mention also some other complications.\\
(1) It will be hard to see the ``singularities'' with staggered fermions
for $\beta<6$. This is because $\widetilde M_{ii} - M_{ii}$ grows
like $a^2$ (at fixed physical quark mass).
Indeed, by $\beta=5.7$ the flavor singlet pion has a 
mass comparable to $M_\rho$!
Thus the $\eta'$ is no longer light, 
and its loop effects will be suppressed.
In fact, the rise in $M_\pi^2/m_q$ as $m_q\to0$ for $\beta=5.7$ is very
gradual\cite{gottlieb96}, and could be due to the $c_2$ term.\\
(2) It will be hard to see the singularities using Wilson fermions.
This is because we do not know, {\it a priori}, where $m_q$ vanishes,
and, as shown by Mawhinney, it is hard to distinguish
the log divergence of QChPT from an offset in $m_q$.\\
(3) A related log divergence is predicted for $\langle\bar\psi\psi\rangle$,
which has not been seen so far \cite{kimsinclair,mawhinney95}.
It is not clear to me that this is a problem for QChPT, however,
because it is difficult to extract the non-perturbative part of
$\langle\bar\psi\psi\rangle$ from the quadratically divergent perturbative
background.

Two other quantities give evidence concerning $\delta$.
The first uses the ratio of decay constants 
\begin{equation}
R_{BG} = f_{12}^2/(f_{11} f_{22}) \,.
\end{equation}
This is designed to cancel the analytic terms proportional 
to $m_q$ \cite{BGI}, leaving a non-analytic term proportional to $\delta$.
The latest analysis finds $\delta\approx 0.14$ \cite{guptafpi}.
It is noteworthy that a good fit once again requires
the inclusion of $O(m_q^2)$ terms.

The second quantity is the double difference
\begin{equation}
{\rm ES}2 = (M_\Omega - M_\Delta) - 3 (M_{\Xi^*} - M_{\Sigma^*}) \,,
\end{equation}
which is one measure of the breaking of the equal spacing rule for decuplets.
This is a good window on artifacts due to quenching because
its expansion begins at $O(M_\pi^5)$ in QCD, but contains terms
proportional to $\delta M_\pi^2 \log M_\pi$ in QQCD \cite{LS}.
The LANL group finds that ${\rm ES}2$ differs from zero 
by 2-$\sigma$ \cite{ourspect96},
and I find that the data can be fit with $\delta\approx 0.1$ 
\cite{sharpebary}.

In my view, the preponderance of the evidence suggests a value of
$\delta$ in the range $0.1-0.2$. All extractions are complicated by
the fact that the effects proportional to $\delta$ are small with
present quark masses. 
To avoid them, one should use quark masses above $m_s/4-m_s/3$.
This is true not only for the light quark quantities
discussed above, but also for heavy-light quantities such as $f_B$.
This, too, is predicted to be singular as the light quark mass vanishes
\cite{zhang}.

\section{QUENCHING ERRORS}
\label{sec:querr}

I close the first part of the talk by listing, in Table \ref{tab:querr},
a sampling of estimates of quenching errors, defined by
\begin{equation}
{\rm Error}({\rm Qty}) = { [ {\rm Qty}({\rm QCD})- {\rm Qty}({\rm QQCD})]
\over{\rm Qty}({\rm QCD})} \,.
\end{equation}
I make the estimates by taking 
the numerator to be the difference between the pion loop contributions
in the full and quenched chiral expansions.
To obtain numerical values I set $\Lambda=m_\rho$ ($\Lambda$ is
the scale occurring in chiral logs), use $f=f^Q=f_K$, and assume
$\delta=0.1$ and $\alpha_\Phi=0$. For the estimates of heavy-light
quantities I set $g'=0$, where $g'$ is an $\eta'$-$B$-$B$ coupling defined
in Ref. \cite{zhang}. These estimates assume that the extrapolation
to the light quark mass is done linearly from $m_q\approx m_s/2$.
For example, $f_{B_d}$ in QQCD is {\em not} the quenched value with the
physical $d$-quark mass (which would contain a large artifact proportional
to $\delta$), but rather the value obtained by linear extrapolation from
$m_s/2$, where the $\delta$ terms are much smaller.
This is an attempt to mimic what is actually done in numerical simulations.

\begin{table}[tb]
\caption{Estimates of quenching errors.}
\label{tab:querr}
\begin{tabular}{ccl}
\hline
Qty.		&Ref.			& Error \\
\hline
$f_\pi/M_\rho$	&\cite{gassleut,sharpechbk}	& $\,\sim 0.1$ \\
$f_K/f_\pi-1$	& \cite{BGI}			& $\ 0.4$ \\
$f_{B_s}$	& \cite{zhang}			& $\ 0.2$\\
$f_{B_s}/f_{B_d}$ & \cite{zhang}		& $\ 0.16$ \\
$B_{B_s}/B_{B_d}$ & \cite{zhang}		& $\,-0.04$ \\
$B_K$ ($m_d=m_s$) & \cite{sharpechbk}		& $\ 0$ \\ 
$B_K$ ($m_d\ne m_s$) & \cite{sharpetasi}		& $\ 0.05$ \\ 
$M_\Xi-M_\Sigma$& \cite{sharpebary}		& $\ 0.4$ \\
$M_\Sigma-M_N$	& \cite{sharpebary}		& $\ 0.3$ \\
$M_\Omega-M_\Delta$& \cite{sharpebary}		& $\ 0.3$ \\
\hline
\end{tabular}
\vspace{-0.2truein}
\end{table}

For the first two estimates, I have used the facts that, in QCD, 
\cite{gassleut}
\begin{eqnarray}
f_\pi &\approx& f\, [1 - 0.5 L(M_K)] 
\,,\\
f_K/f_\pi &\approx& 1 - 0.25 L(M_K) - 0.375 L(M_\eta) 
\,,
\end{eqnarray}
(where $L(M) = (M/4\pi f)^2 \log(M^2/\Lambda^2)$, and $f_\pi= 93\,$MeV),
while in QQCD \cite{BGI,sharpechbk} 
\begin{eqnarray}
f_\pi &\approx& f^Q \,,\\
{f_K\over f_\pi} &\approx& 1 + {\delta \over 2} \left[
{M_K^2 \over M_{ss}^2 - M_\pi^2} \log {M_{ss}^2 \over M_\pi^2} - 1 \right]
\,.
\end{eqnarray}
I have not included the difference of pion loop contributions to $M_\rho$,
since the loop has not been evaluated in QChPT, 
and a model calculation suggests that the difference is small 
\cite{cohenleinweber}.
Details of the remaining estimates can be found in the references.

Let me stress that these are estimates and not calculations.
What they give is a sense of the effect of quenching on
the contributions of ``pion'' clouds surrounding hadrons---these clouds are
very different in QQCD and QCD!
But this difference in clouds could be cancelled numerically by differences
in the analytic terms in the chiral expansion.
As discussed in Ref. \cite{zhang}, a more conservative view is thus to
treat the estimates as rough upper bounds on the quenching error.
Those involving ratios (e.g. $f_K/f_\pi$)
are probably more reliable since some of the analytic terms do not contribute.
One can also form double ratios for which the error 
estimates are yet more reliable
(e.g. $R_{BG}$ and ES2 from the previous section; see also Ref. \cite{zhang}), 
but these quantities are of less phenomenological interest.

My aim in making these estimates is to obtain a sense of which
quenched quantities are likely to be more reliable and which less,
and to get an sense of the possible size of quenching errors.
My conclusion is that the errors could be significant 
in a number of quantities, including those involving heavy-light mesons.
One might have hoped that the ratio $f_{B_s}/f_{B_d}$ would have
small quenching errors, but the chiral loops indicate otherwise.
For some other quantities, such as $B_{B_s}/B_{B_d}$ and $B_K$, 
the quenching errors are likely to be smaller.

If these estimates work, then it will be worthwhile extending them
to other matrix elements of phenomenological interest, e.g.
$K\to\pi\pi$ amplitudes.
Then, when numerical results in QQCD are obtained,
we have at least a rough estimate of the quenching error in hand.
Do the estimates work? As we will see below,
those for $f_\pi/m_\rho$, $f_K/f_\pi$ and $B_K$ are consistent
with the numerical results obtained to date.

\section{RESULTS FOR DECAY CONSTANTS}
\label{sec:decayc}
\begin{figure}[tb]
\vspace{-0.6truein}
\centerline{\psfig{file=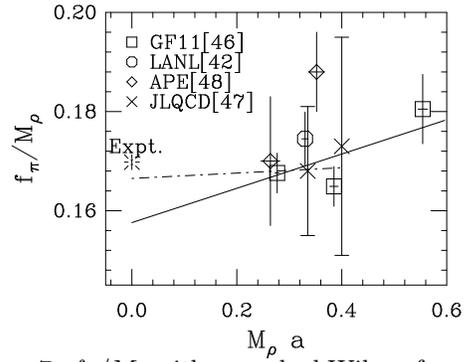,height=2.5truein}}
\vspace{-0.6truein}
\caption{$f_\pi/M_\rho$ with quenched Wilson fermions.}
\vspace{-0.2truein}
\label{fig:fpi_mrhoW}
\end{figure}

\begin{figure}[tb]
\vspace{-0.1truein}
\centerline{\psfig{file=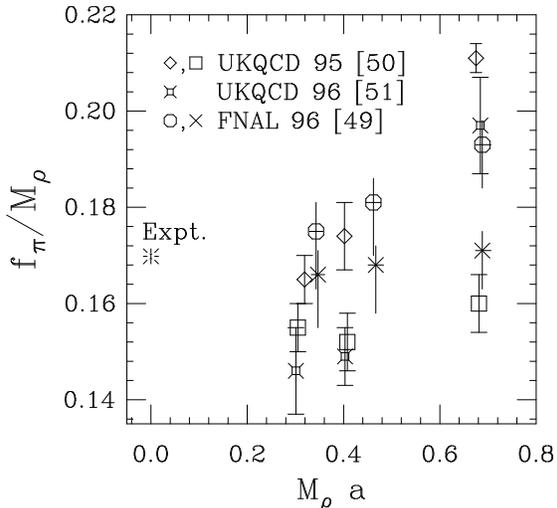,height=3.truein}}
\vspace{-0.6truein}
\caption{$f_\pi/M_\rho$ with quenched SW fermions.}
\vspace{-0.2truein}
\label{fig:fpi_mrhoCL}
\end{figure}

For the remainder of the talk I will don the hat of a reviewer,
and discuss the status of results for weak matrix elements.
All results will be quenched, unless otherwise noted.
I begin with $f_\pi/M_\rho$, the results for which are shown
in Figs. \ref{fig:fpi_mrhoW} (Wilson fermions) and \ref{fig:fpi_mrhoCL}
(SW fermions, with tadpole improved $c_{SW}$).
The normalization here is $f_\pi^{\rm expt}=0.131\,$MeV,
whereas I use $93\,$MeV elsewhere in this talk.

Consider the Wilson data first. One expects a linear dependence on $a$,
and the two lines are linear extrapolations taken from Ref. \cite{guptafpi}.
The solid line is a fit to all the data, 
while the dashed curve excludes the point with largest $a$
(which might lie outside the linear region).
It appears that the quenched result is lower than experiment,
but there is a $5-10\%$ uncertainty.
Improving the fermion action (Fig. \ref{fig:fpi_mrhoCL}) 
doesn't help much because of
uncertainties in the normalization of the axial current.
For the FNAL data, the upper (lower) points correspond to
using $\alpha_s(\pi/a)$ ($\alpha_s(1/a)$) in the matching factor.
The two sets of UKQCD95 points correspond to different normalization schemes.
Again the results appear to extrapolate to a point below experiment.

It is disappointing that we have not done better with such a basic quantity.
We need to reduce both statistical errors and normalization uncertainty.
The latter may require non-perturbative methods, or the use of
staggered fermions (where $Z_A=1$).
Note that chiral loops estimate that the quenched result 
will undershoot by 12\%, 
and this appears correct in sign, and not far off in magnitude.

\begin{figure}[tb]
\vspace{-0.1truein}
\centerline{\psfig{file=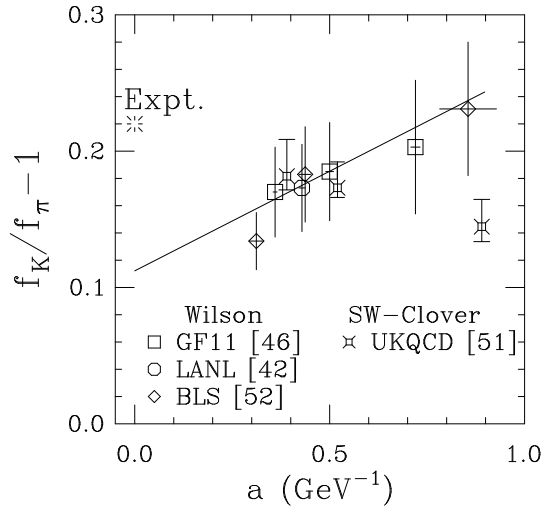,height=3.0truein}}
\vspace{-0.6truein}
\caption{$(f_K-f_\pi)/f_\pi$ in quenched QCD.}
\vspace{-0.2truein}
\label{fig:fk_fpi}
\end{figure}

Results for $(f_K-f_\pi)/f_\pi$ are shown in Fig. \ref{fig:fk_fpi}.
This ratio measures the mass dependence of decay constants.
Chiral loops suggest a 40\% underestimate in QQCD.
The line is a fit to all the Wilson data (including the largest $a$'s),
and indeed gives a result about half of the experimental value.
The new UKQCD results, using tadpole improved SW fermions,
are, by contrast, rising towards the experimental value.
It will take a substantial reduction in statistical errors to sort this out.

\section{STATUS OF $B_K$: STAGGERED}
\label{sec:bks}

$B_K$ is defined by
\begin{equation}
B_K = 
{\langle \bar K| \bar s \gamma_\mu^L d\, \bar s\gamma_\mu^L d | K \rangle 
 \over
(8/3) \langle \bar K| \bar s \gamma_\mu^L d|0 \rangle
      \langle 0 |\bar s\gamma_\mu^L d | K \rangle } \,.
\label{eq:bkdef}
\end{equation}
It is a scale dependent quantity, and I will quote results in
the NDR (naive dimensional regularization) scheme at 2 GeV.
It can be calculated with very small statistical errors,
and has turned out to be a fount of knowledge about systematic errors.
This is true for both staggered and Wilson fermions,
though for different reasons.

There has been considerable progress with both types of fermions
in the last year. I begin with staggered fermions,
which hold the advantage for $B_K$ as they have a remnant chiral symmetry. 
Back in 1989, I thought we knew what the 
quenched answer was, based on calculations at $\beta=6$ on 
$16^3$ and $24^3$ lattices:
$B_K = 0.70(2)$ \cite{sharpe89,ourbkprl}.
I also argued that quenching errors
were likely small (see Table \ref{tab:querr}).
I was wrong on the former, though maybe not on the latter.

By 1993, Gupta, Kilcup and I had found that $B_K$ had a
considerable $a$ dependence \cite{sharpe93}.
Applying Symanzik's improvement program, I argued that the discretization
errors in $B_K$ should be $O(a^2)$, and not $O(a)$.
Based on this, we extrapolated our data quadratically, and quoted
$B_K(NDR,2{\rm GeV}) = 0.616(20)(27)$ for the quenched result.
Our data alone, however, was not good enough to distinguish linear and
quadratic dependences.

Last year, JLQCD presented results from a more
extensive study (using $\beta=5.85$, $5.93$, $6$ and $6.2$) \cite{jlqcdbk95}.
Their data strongly favored a linear dependence on $a$. 
If correct, this would lead to a value of $B_K$ close to $0.5$.
The only hope for someone convinced of an $a^2$ dependence was 
competition between a number of terms.

\begin{figure}[tb]
\vspace{-0.1truein}
\centerline{\psfig{file=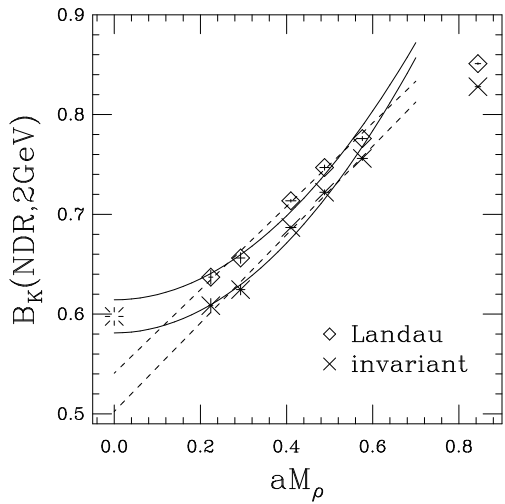,height=3.0truein}}
\vspace{-0.6truein}
\caption{JLQCD results for staggered $B_K$.}
\vspace{-0.2truein}
\label{fig:jlqcdbk}
\end{figure}

Faced with this contradiction between numerical data and theory,
JLQCD have done further work on both fronts \cite{aoki96}. 
They have added two additional lattice spacings, $\beta=5.7$ and $6.4$,
thus increasing the lever arm. They have also carried out finite volume
studies at $\beta=6$ and $6.4$, finding only a small effect.
Their data are shown in Fig. \ref{fig:jlqcdbk}. 
``Invariant'' and ``Landau'' refer to two possible discretizations
of the operators---the staggered fermion operators are spread out over
a $2^4$ hypercube, and one can either make them gauge invariant
by including gauge links, or by fixing to Landau gauge and omitting the links.
The solid (dashed) lines show quadratic (linear) fits to the first five
points. The $\chi^2/{\rm d.o.f.}$ are
\begin{center}
\begin{tabular}{ccc}
\hline
Fit 	& Invariant 	& Landau \\
$a$	& 0.86		& 0.67	\\
$a^2$	& 1.80		& 2.21	\\
\hline
\end{tabular}
\end{center}
thus favoring the linear fit, but by a much smaller difference than
last year. If one uses only the first four points then linear
and quadratic fits are equally good.
What has changed since last year is that the new point at $\beta=6.4$
lies above the straight line passing through the next four points.

JLQCD have also checked the theoretical argument using a simpler method
of operator enumeration\cite{aoki96,ishizuka96}.\footnote{
A similar method has also been introduced by Luo \cite{luo96}.}
The conclusion is that there cannot be $O(a)$ corrections to $B_K$, 
because there are no operators available with which one
could remove these corrections.
Thus JLQCD use quadratic extrapolation and quote
(for degenerate quarks)
\begin{equation}
B_K({\rm NDR}, 2\,{\rm GeV}) = 0.5977 \pm 0.0064 
\pm 0.0166 \,,
\label{eq:jlqcdbk}
\end{equation}
where the first error is statistical, the second due to truncation of
perturbation theory.
This new result agrees with that from 1993 (indeed, the results are
consistent at each $\beta$), but has much smaller errors.
To give an indication of how far things have come, compare our 
1993 result at $\beta=6$  with Landau-gauge operators, $0.723(87)$ 
\cite{sharpe93}, to the corresponding JLQCD result $0.714(12)$.

The perturbative error in $B_K$ arises from truncating the
matching of lattice and continuum operators to one-loop order.
The use of two different lattice operators allows one to estimate this
error without resort to guesswork about the higher order terms in the
perturbative expansion.
The difference between the results from the two operators is
of $O[\alpha(2\,{\rm GeV})^2]$, and thus should remain
finite in the continuum limit.
This is what is observed in Fig. \ref{fig:jlqcdbk}.

I will take Eq. \ref{eq:jlqcdbk} as the best estimate of $B_K$ in QQCD.
The errors are so much smaller than those in previous staggered results
and in the results with Wilson fermions discussed below, 
that the global average is 
not significantly different from the JLQCD number alone.
The saga is not quite over, however, since one should 
confirm the $a^2$ dependence by running at even smaller lattice spacings.
JLQCD intend to run at $\beta=6.6$.
If the present extrapolation holds up, then it shows how one must
beware of keeping only a single term when extrapolating in $a$.

\subsection{Unquenching $B_K$}

To obtain a result for QCD proper, two steps remain:
the inclusion of dynamical quarks, and the use of $m_s\ne m_d$.
The OSU group has made important progress on the first step \cite{osubk96}.
Previous studies (summarized in Ref. \cite{soni95}) found that
$B_K$ was reduced slightly by sea quarks, although the effect was
not statistically significant.
The OSU study, by contrast, finds a statistically significant
increase in $B_K$
\begin{equation}
{ B_K({\rm NDR,2\,GeV},N_f=3) \over B_K({\rm NDR,2\,GeV},N_f=0)} 
= 1.05 \pm 0.02 \,.
\label{eq:bkquerr}
\end{equation}
They have improved upon previous work by reducing statistical errors,
and by choosing their point at $N_f=0$ ($\beta=6.05$) 
to better match the lattice spacing at $N_f=2$ ($\beta=5.7$, $m_qa=0.01$)
and $4$ ($\beta=5.4$, $m_qa=0.01$).

There are systematic errors in this result
which have yet to be estimated.
First, the dynamical lattices are chosen to have 
$m_q^{\rm sea}=m_q^{\rm val} = m_s^{\rm phys}/2$,
and so they are truely unquenched simulations.
But $m_s^{\rm phys}$ is determined by 
extrapolating in the valence quark mass alone,
and is thus a partially quenched result.
This introduces an uncertainty in $m_s$ which feeds into the estimate of the
$N_f$ dependence of $B_K$.
Similarly, $a$ is determined by a partially quenched extrapolation, 
resulting in an uncertainty in the
matching factors between lattice and continuum operators.

But probably the most important error comes from the possibility
of significant $a$ dependence in the ratio in Eq. \ref{eq:bkquerr}. 
The result quoted is for $a^{-1}=2\,$GeV, at which $a$ the
discretization error in the quenched $B_K$ is 15\%.
It is not inconceivable that, say, $B_K$ in QCD has very little dependence
on $a$, in which case the ratio would increase to $\sim 1.2$ in the
continuum limit.
Clearly it is very important to repeat the comparison at a different
lattice spacing.

Despite these uncertainties, I will take the OSU result and error
as the best estimate of the effect of quenching at $a=0$.
I am being less conservative than I might be because a small
quenching error in $B_K$ is consistent with the expectations of QChPT.
A more conservative estimate for the ratio would be $1.05\pm0.15$.

\subsection{$B_K$ for non-degenerate quarks}
\label{subsec:bknondegen}

What remains is to extrapolate from $m_s=m_d\approx m_s^{\rm phys}/2$ 
to $m_s=m_s^{\rm phys}$ and $m_d=m_d^{\rm phys}$.
This appears difficult because it requires
dynamical quarks with very small masses.
This may not be necessary, however, if one uses ChPT to guide
the extrapolation \cite{sharpetasi}.
The point is that the chiral expansion in QCD is \cite{bijnens,sharpechbk}
\begin{equation}
{B_K\over B} = 1 - \left(3+{\epsilon^2 \over 3}\right) y\ln{y}
		+ b y + c y \epsilon^2 \,,
\label{eq:bkchqcd}
\end{equation}
where
\begin{equation}
\epsilon=(m_s-m_d)/(m_s+m_d)\,,\ 
y = M_K^2/(4 \pi f)^2 ,
\end{equation}
and $B$, $b$ and $c$ are unknown constants.
At this order $f$ can be equally well taken to be $f_\pi$ or $f_K$.
Equation \ref{eq:bkchqcd}
is an expansion in $y$, but is valid for all $\epsilon$.
The idea is to determine $c$ by working at small $\epsilon$, 
and then use the formula to extrapolate to $\epsilon=1$.
This ignores corrections of $O(y^2)$, and so the errors
in the extrapolation are likely to be $\sim 25\%$.

Notice that $m_u$ does not enter into Eq. \ref{eq:bkchqcd}.
Thus one can get away with a simulation using only two
different dynamical quark masses, e.g. setting
$m_u=m_d < m_s^{\rm phys}/2$, while holding $m_s +m_d = m_s^{\rm phys}$.
To date, no such calculation has been done.
To make an estimate I use the chiral log alone, i.e. set $c=0$, yielding
\begin{equation}
B_K({\rm non-degen}) = (1.04-1.08) B_K({\rm degen}) \,.
\end{equation}
The range comes from using $f=f_\pi$ and $f_K$, and varying the
scale in the logarithm from $m_\rho-1\,$GeV.
Since the chiral log comes mainly from kaon 
and $\eta$ loops \cite{sharpechbk}, I prefer $f=f_K$, which leads to
$1.04-1.05$ for the ratio.
To be conservative I take $1.05\pm0.05$, and assume that the generous
error is large enough to include also the error in the estimate
of the effect of unquenching. This leads to a final estimate of
\begin{equation}
B_K({\rm NDR},{\rm 2\,GeV,QCD}) = 0.66 \pm 0.02 \pm 0.03 \,,
\label{eq:finalqcdbk}
\end{equation}
where the first error is that in the quenched value, the second that
in the estimate of unquenching and using non-degenerate quarks.
Taking the more conservative estimate of the unquenching error (15\%),
and adding it in quadrature with the (5\%) estimate of the error
in accounting for non-degenerate quarks, increases the second error
in Eq. \ref{eq:finalqcdbk} to $0.11$.

It is customary to quote a result for the renormalization group
invariant quantity
\[
{\widehat{B}_K \over B_K(\mu)} = \alpha_s(\mu)^{-\gamma_0 \over2\beta_0}
\left(1 + 
{\alpha_s(\mu) \over 4 \pi} \left[{\beta_1 \gamma_0 -\beta_0\gamma_1
\over 2 \beta_0^2} \right] \right)
\]
in the notation of Ref. \cite{crisafulli}.
Using $\alpha_s(2\,{\rm GeV})=0.3$ and $N_f=3$, I find
$\widehat{B}_K=0.90(3)(4)$, with the last error increasing to
$0.14$ with the more conservative error. 
This differs from the result I quoted in Ref. \cite{sharpe93}, 
because I am here using the 2-loop formula
and a continuum choice of $\alpha_s$. 

\subsection{Chiral behavior of $B_K$}
\label{subsec:bkchfit}

Since $B_K$ can be calculated very accurately, it provides a 
potential testing ground for (partially) quenched ChPT. 
This year, for the first time, such tests have been undertaken,
with results from OSU \cite{osubk96}, 
JLQCD \cite{aoki96}, and Lee and Klomfass \cite{lee96}.
I note only some highlights.

It turns out that, for $\epsilon=0$,
Eq. \ref{eq:bkchqcd} is valid for all $N_f$ \cite{sharpechbk}.
This is why my estimate of the quenching error for $B_K$ with degenerate
quarks in Table \ref{tab:querr} is zero.
Thus the first test of (P)QChPT is to see
whether the $-3 y\ln y$ term is present.
The OSU group has the most extensive data as a function of $y$,
and indeed observe curvature of the expected sign and magnitude for 
$N_f=0,2,4$.
JLQCD also finds reasonable fits to the chiral form, 
as long as they allow a substantial dependence of $f$ on lattice spacing.
They also study other $B$ parameters, with similar conclusions.

Not everything works. JLQCD finds that the volume dependence predicted
by the chiral log \cite{sharpechbk} is too small to fit their data.
Fitting to the expected form for $\epsilon\ne0$ in QQCD, they find
$\delta=-0.3(3)$, i.e. of the opposite sign to the other determinations
discussed in Sec. \ref{sec:chevidence}.
Lee and Klomfass have studied the $\epsilon$ dependence with $N_f=2$
(for which there is as yet no PQChPT prediction).
It will be interesting to see how things evolve.
My only comment is that one may need to include $O(y^2)$ terms in the
chiral fits.

\section{STATUS OF $B_K$: WILSON}
\label{sec:bkw}

There has also been considerable progress in the last year in the
calculation of $B_K$ using Wilson and SW fermions.
The challenge here is to account for the effects of the
explicit chiral symmetry breaking in the fermion action.
Success with $B_K$ would give one confidence to attempt
more complicated calculations.

The operator of interest,
\begin{equation}
{\cal O}_{V+A} = \bar s \gamma_\mu d\, \bar s \gamma_\mu d +
	 \bar s \gamma_\mu \gamma_5 d\, \bar s \gamma_\mu \gamma_5 d
\,,
\end{equation}
can ``mix'' with four other dimension 6 operators
\begin{equation}
{\cal O}_{V+A}^{\rm cont} = Z_{V+A} \left(
{\cal O}_{V+A} + \sum_{i=1}^{4} z_i {\cal O}_i \right) + O(a) 
\end{equation}
where the ${\cal O}$ on the r.h.s. are lattice operators.
The ${\cal O}_i$ are listed in Refs. \cite{kuramashi96,talevi96}.
The meaning of this equation is that, 
for the appropriate choices of $Z_{V+A}$ and the $z_i$, 
the lattice and continuum operators will have the same
matrix elements, up to corrections of $O(a)$.
In particular, while the matrix elements of a general four fermion operator
has the chiral expansion
\begin{eqnarray}
\lefteqn{\langle \bar K | {\cal O} | K \rangle =
\alpha + \beta M_K^2 + \delta_1 M_K^4 +} \\
&& p_{\bar K}\cdot p_{K} 
(\gamma + \delta_2 M_K^2 + \delta_3 p_{\bar K}\cdot p_{K} ) +\dots \,,
\end{eqnarray}
chiral symmetry implies that $\alpha=\beta=\delta_1=0$ for 
the particular operator ${\cal O}={\cal O}_{V+A}^{\rm cont}$.
Thus, one can test that the $z_i$ are correct by checking that the
first three terms are absent.\footnote{I have ignored chiral logarithms,
which will complicate the analysis, but can probably be ignored given
present errors and ranges of $M_K$.}
Note that the $z_i$ must be known quite accurately because the
terms we are removing are higher order in the chiral 
expansion than the terms we are keeping.

Five methods have been used to determine the $z_i$ and $Z_{V+A}$. \\
(1)
One-loop perturbation theory. This fails to give the correct
chiral behavior, even when tadpole improved. \\
(2)
Use (1) plus enforce chiral behavior by adjusting subsets of the
$z_i$ by hand \cite{bernardsoni89}. Different subsets give differing
results, introducing an additional systematic error. 
Results for a variety of $a$ were presented by 
Soni last year \cite{soni95}.\\
(3)
Use (1) and discard the $\alpha$, $\beta$ and $\gamma$ terms,
determined by doing the calculation at a variety of momenta. 
New results using this method come from the LANL group \cite{gupta96}.
Since the $z_i$ are incorrect, there is, however, 
an error of $O(g^4)$ in $B_K$. \\
(4)
Non-perturbative matching by imposing continuum normalization conditions
on Landau-gauge quark matrix elements. This approach has been pioneered
by the Rome group, and is reviewed here by Rossi \cite{rossi96}.
The original calculation omitted one operator \cite{donini}, 
but has now been corrected \cite{talevi96}. \\
(5)
Determine the $z_i$ non-perturbatively by
imposing chiral ward identities on quark matrix elements.
Determine $Z_{V+A}$ as in (4).
This method has been introduced by JLQCD \cite{kuramashi96}. 

The methods of choice are clearly (4) and (5), 
as long as they can determine the $z_i$ accurately enough.
In fact, both methods work well:
the errors in the non-perturbative results are much smaller 
than their difference from the one-loop perturbative values.
And both methods find that the matrix element of ${\cal O}_{V+A}^{\rm cont}$
has the correct chiral behavior, within statistical errors.
What remains to be studied is the uncertainty introduced by the fact that
there are Gribov copies in Landau gauge. Prior experience suggests
that this will be a small effect.

It is not yet clear which, if either, of methods (4) and (5) is preferable
for determining the $z_i$.
As stressed in Ref. \cite{talevi96} the $z_i$ are unique,
up to corrections of $O(a)$.
In this sense, both methods must give the same results.
But they are quite different in detail,
and it may be that the errors are smaller with one method or the other. 
It will be interesting to see a comprehensive comparison between them
and also with perturbation theory.

\begin{figure}[tb]
\vspace{-0.1truein}
\centerline{\psfig{file=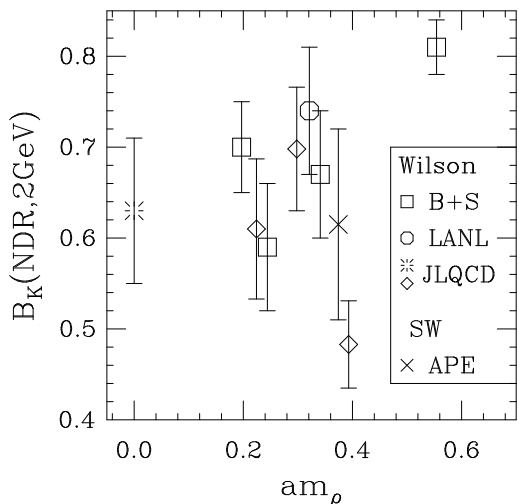,height=3.0truein}}
\vspace{-0.6truein}
\caption{Quenched $B_K$ with Wilson fermions.}
\vspace{-0.2truein}
\label{fig:bkw}
\end{figure}

In Fig. \ref{fig:bkw} I collect the results for $B_K$. 
All calculations use $m_s=m_d$ and the quenched approximation.
The fact that most of the results agree is a significant success, 
given the variety of methods employed.
It is hard to judge which method gives the smallest errors,
because each group uses different ensembles and lattice sizes,
and estimates systematic errors differently.
The errors are larger than with staggered fermions mostly because
of the errors in the $z_i$.

Extrapolating to $a=0$ using the data in Fig. \ref{fig:bkw} would
give a result with a large uncertainty.
Fortunately, JLQCD has found a more accurate approach.
Instead of $B_K$, they consider the ratio of the matrix element
of ${\cal O}_{V+A}^{\rm cont}$ to its vacuum saturation approximant.
The latter differs from the denominator of $B_K$ (Eq. \ref{eq:bkdef})
at finite lattice spacing. The advantage of this choice is
that the $z_i$ appear in both the numerator and denominator,
leading to smaller statistical errors.
The disadvantage is that the new ratio has the wrong chiral behavior at
finite $a$. 
It turns out that there is an overall gain, and from their
calculations at $\beta=5.9$, $6.1$ and $6.3$ they find
$B_K({\rm NDR,2\,GeV})=0.63(8)$. 
This is the result shown at $a=0$ in Fig. \ref{fig:bkw}.
It agrees with the staggered result, although it has much larger errors.
Nevertheless, it is an important consistency check,
and is close to ruling out the use of a linear extrapolation
in $a$ with staggered fermions.

\section{OTHER MATRIX ELEMENTS}
\label{sec:otherme}

The LANL group \cite{gupta96} has quenched results (at $\beta=6$)
for the matrix elements which determine the dominant part 
of the electromagnetic penguin contribution to $\epsilon'/\epsilon$
\begin{eqnarray}
 B_7^{I=3/2} &=&  0.58 \pm 0.02 {\rm (stat)} {+0.07 \atop -0.03} \,, \\
 B_8^{I=3/2} &=&  0.81 \pm 0.03 {\rm (stat)} {+0.03 \atop -0.02} \,.
\end{eqnarray}
These are in the NDR scheme at 2 GeV.
The second error is from the truncation of perturbation matching factors.
These numbers lie at or below the lower end of the range used by
phenomenologists.
The LANL group also finds $B_D=0.78(1)$.

There are also new results for $f_\rho$ and $f_\phi$ 
\cite{guptafpi,yoshie96}, for the pion polarizability \cite{wilcox96},
and for strange quark contributions to magnetic moments \cite{dong96}.

\section{FUTURE DIRECTIONS}
\label{sec:future}

This year has seen the first detailed tests of the predicted
chiral behavior of quenched quantities.
Further work along these lines will help us make better extrapolations,
and improve our understanding of quenching errors.
It is also a warm-up exercise for the use of chiral perturbation theory
in unquenched theories. I have outlined one such application in
Sec. \ref{subsec:bknondegen}. I expect the technique to be
of wide utility given the difficulty in simulating light dynamical
fermions.

As for matrix elements, there has been substantial progress on $B_K$.
It appears that we finally know the quenched result, 
thanks largely to the efforts of JLQCD.
At the same time, it is disturbing that the complicated $a$ dependence
has made it so difficult to remove the last 20\% of the errors.
One wonders whether similar complications lie lurking beneath
the relatively large errors in other matrix elements.

The improved results for $B_K$ with Wilson fermions show that
non-perturbative normalization of operators is viable.
My hope is that we can now return to an issue 
set aside in 1989: the calculation of $K\to\pi\pi$ amplitudes.
The main stumbling block is the need
to subtract lower-dimension operators.
A method exists for staggered fermions, but the errors have so far
swamped the signal. 
With Wilson fermions, one needs a non-perturbative method, and the hope
is that using quark matrix elements in Landau gauge
will do the job \cite{talevi96}.
Work is underway with both types of fermion.
Given the success of the Schr\"odinger functional method at calculating
current renormalizations \cite{wittig96},
it should be tried also for four fermion operators.

Back in 1989, I also described preliminary work
on non-leptonic $D$ decays, e.g. $D\to K\pi$. 
Almost no progress has been made since then,
largely because we have been lacking
a good model of the decay amplitude for Euclidean momenta.
A recent proposal by Ciuchini {\em et al.} may fill this gap \cite{ciuchini}.

Enormous computational resources have been used to calculate 
matrix elements in (P)QQCD.
To proceed to QCD at anything other than a snail's pace may well
require the use of improved actions.
Indeed, the large discretization errors in quenched staggered $B_K$
already cry out for improvement.
The fact that we know $B_K$ very accurately
will provide an excellent benchmark for such calculations.
Working at smaller values of the cut-off, $1/a$,
alleviates some problems while making others worse.
Subtraction of lower dimension operators becomes simpler,
but the evaluation of mixing with operators of the same dimension
becomes more difficult. 
It will be very interesting to see how things develop.

\section*{Acknowledgements}
I am grateful to Peter Lepage for helpful conversations 
and comments on the manuscript.

%

\def\PRL#1#2#3{{Phys. Rev. Lett.} {\bf #1}, #3 (#2)}
\def\PRD#1#2#3{{Phys. Rev.} {\bf D#1}, #3 (#2)}
\def\PLB#1#2#3{{Phys. Lett.} {\bf #1B} (#2) #3}
\def\NPB#1#2#3{{Nucl. Phys.} {\bf B#1} (#2) #3}
\def\NPBPS#1#2#3{{Nucl. Phys.} {\bf B({Proc.Suppl.}) {#1}} (#2) #3}
\def\etal{{\em et al.}}

\end{document}